\documentclass{ecai} 

\usepackage{latexsym}
\usepackage{amssymb}
\usepackage{amsmath}
\usepackage{amsthm}
\usepackage{booktabs}
\usepackage{enumitem}
\usepackage{graphicx}
\usepackage{color}
\usepackage{multirow}
\usepackage{booktabs}
\usepackage{makecell}
\usepackage[table]{xcolor}

\newcommand{\BibTeX}{B\kern-.05em{\sc i\kern-.025em b}\kern-.08em\TeX}

\begin{document}


\begin{frontmatter}


\paperid{5031} 


\title{A Fast and Lightweight Model for Causal Audio-Visual Speech Separation}


\author[A,B,C]{\fnms{}~\snm{Wendi Sang}\footnote{Equal contribution (co-first authors).}}
\author[C]{\fnms{}~\snm{Kai Li}\footnotemark[1]}
\author[C]{\fnms{}~\snm{Runxuan Yang}}
\author[A,B]{\fnms{}~\snm{Jianqiang Huang}\thanks{Co-corresponding author. Email: hjqxaly@163.com, xlhu@tsinghua.edu.cn}}
\author[C,D]{\fnms{}~\snm{Xiaolin Hu}\footnotemark[*]}

\address[A]{School of Computer Technology and Application, Qinghai University, Xining 810016, China}
\address[B]{Intelligent Computing and Application Laboratory of Qinghai Province, Qinghai University, Xining 810016, China}
\address[C]{Department of Computer Science and Technology, Institute for AI,  \\ BNRist, Tsinghua Laboratory of Brain and Intelligence (THBI), \\ IDG/McGovern Institute for Brain Research, Tsinghua University, Beijing 100084, China}
\address[D]{Chinese Institute for Brain Research (CIBR), Beijing 100010, China}


\begin{abstract}
Audio-visual speech separation (AVSS) aims to extract a target speech signal from a mixed signal by leveraging both auditory and visual (lip movement) cues. However, most existing AVSS methods exhibit complex architectures and rely on future context, operating offline, which renders them unsuitable for real-time applications. Inspired by the pipeline of RTFSNet, we propose a novel streaming AVSS model, named Swift-Net, which enhances the causal processing capabilities required for real-time applications. Swift-Net adopts a lightweight visual feature extraction module and an efficient fusion module for audio-visual integration. Additionally, Swift-Net employs Grouped SRUs to integrate historical information across different feature spaces, thereby improving the utilization efficiency of historical information. We further propose a causal transformation template to facilitate the conversion of non-causal AVSS models into causal counterparts. Experiments on three standard benchmark datasets (LRS2, LRS3, and VoxCeleb2) demonstrated that under causal conditions, our proposed Swift-Net exhibited outstanding performance, highlighting the potential of this method for processing speech in complex environments.
\end{abstract}

\end{frontmatter}


\section{Introduction}

With the rapid advancement of speech processing and artificial intelligence technologies, audio-only speech separation (AOSS) methods are facing numerous challenges in the cocktail party scenario \cite{michelsanti2021overview}. These challenges are mainly manifested as a significant decline in processing efficiency with an increasing number of speakers, and insufficient robustness in complex noisy environments \cite{rahimi2022reading,ephrat2018looking}. Inspired by the human multisensory information processing mechanism \cite{mesgarani2012selective}, researchers have proposed strategies for integrating visual information with auditory information. By incorporating visual cues into the separation network, the robustness and processing efficiency of speech separation methods can be improved \cite{li2024audio,li2023iianet}; Such methods are referred to as audio-visual speech separation (AVSS). 

However, most existing AVSS methods are non-causal and require the entire audio sequence as input, and their complex architectures incur large Params and MACs, limiting them to offline scenarios due to their inability to process streaming data in real-time \cite{liu2020causal}. Therefore, developing low-latency causal models for real-time applications is highly demanded.

In the domain of AOSS, many effective causal methods have already been proposed \cite{luo2019conv, li2022skim,della2024resource}. Unfortunately, straightforward adaptation of these advanced causal AOSS models for AVSS tasks may not yield optimal results. For example, Zhang et al.~\cite{zhang2021online} transformed a causal Conv-TasNet \cite{luo2019conv} into a audio-visual causal version (Causal AV-ConvTasNet), yet the performance was suboptimal (see Section \ref{sec:results}). A natural idea is to construct a causal AVSS system by “causalizing” existing high-performance non-causal backbone networks—simply replacing non-causal CNNs with causal convolutions, replacing bidirectional RNNs with unidirectional ones, using attention mask in Transformers, or converting global pooling into frame-wise pooling (see Section \ref{sec:causal}). To investigate these approaches, we implemented causal variants of several non-causal AVSS models \cite{AV-TFGridNet,lin2023av,li2024audio,tao2024audio,pegg2023rtfs} under the same training and evaluation protocol. Our experimental results revealed that such straightforward causal modifications failed to yield truly lightweight and real-time-supporting models (see Section 5.3). The primary challenge lies in designing a lightweight architecture that can effectively leverage historical information, process audio in a causal manner, and efficiently integrate visual cues.

During this exploration, we did find a non-causal AVSS model, namely RTFSNet \cite{pegg2023rtfs}, that had potential to be converted to an efficient causal model since its overall pipeline is simple and lightweight. However, a straightforward conversion did not lead to satisfactory results (see Section 5.3). After careful analysis, we found that the reason lies in its complex architectures of its visual encoder, separator, and audio-visual fusion network, all of which are constructed using complex architectures such as TDANet \cite{li2022efficient}, convolutional layers, Transformers \cite{vaswani2017attention}, and global pooling operations. These components heavily rely on non-causal structures. When they are directly modified to operate in a causal manner, they inherit the same limitations—namely, restricted access to future context and inefficient modeling of historical information—which results in a substantial degradation of performance.

To address the aforementioned issues, we propose a causal AVSS model, named Swift-Net, building on the RTFSNet's pipeline \cite{pegg2023rtfs}. To reduce computational cost, we adopt SRU networks \cite{lei2018sru} for visual feature extraction (LightVid block) and employ a selective attention mechanism (SAF block) to integrate audio-visual features. To maintain lightweight efficiency while enabling the SRU to better exploit long-range temporal dependencies from historical information by focusing on the temporal context within localized feature channels, we introduce an efficient Grouped SRU mechanism inspired by grouped convolutional layer \cite{krizhevsky2012imagenet} in the separation network (FTGS block). By partitioning feature channels into parallel SRU groups, the model is able to integrate historical information across different feature spaces. This design captures temporal and spectral dependencies with fewer parameters and lower inference latency. In summary, our main contributions are as follows:

\begin{itemize}

    \item We design a lightweight visual feature extraction module that integrates convolution with SRUs to effectively capture historical information.

    \item We propose a lightweight audio-visual fusion module that efficiently integrates auditory features correlated with visual cues through a selective attention mechanism.
    
     \item We design an efficient Grouped SRU module that aggregates historical information across multiple feature spaces through a grouping strategy, further improving model efficiency.

\end{itemize}

Extensive experiments on mainstream datasets LRS2 \cite{afouras2018deep}, LRS3 \cite{afouras2018lrs3}, and VoxCeleb2 \cite{chung2018voxceleb2} demonstrated that Swift-Net achieved SOTA performance. Additionally, we provided a toolkit that modularizes the conversion of mainstream non-causal AVSS methods into causal methods. Source code was made available at \url{https://github.com/JusperLee/Swift-Net}.


\section{Related Work}

\subsection{Non-causal AVSS Model}

Early research on speech separation and enhancement largely centered on audio-only processing \cite{wang2018voicefilter,delcroix2018single,luo2019conv,agrawal2023monaural,ke2023single}. Many of these methods require pre-registered reference utterances to extract the target speaker, which limits their applicability in open-world scenarios. To mitigate these limitations, subsequent studies fused audio with visual cues, giving rise to audio-visual speech separation (AVSS). However, most AVSS models with decent separation performance have adopted non-causal architectures \cite{wu2019time,li2023iianet,li2024audio,afouras2018deep,gao2021visualvoice,lin2023av,tao2024audio,AV-TFGridNet,pegg2023rtfs}, achieving notable improvements in separation performance by integrating both historical and future information. Researchers have developed a variety of model architectures tackling the AVSS task, including convolutional neural networks (CNNs), recurrent neural networks (RNNs), Transformers, and hybrid ones. Specifically, CNN-based models (e.g., AV-ConvTasNet \cite{wu2019time} and CTCNet \cite{li2024audio}) typically utilize time-domain convolutional networks and incorporate visual lip-reading features to enhance separation performance. However, such methods primarily perform audio-visual fusion through convolutional operations, resulting in large model parameter count and high computational complexity. Furthermore, they rely on future time-step information, making them challenging to deploy in real-time processing scenarios. RNN-based models (e.g., AV-DPRNN \cite{tao2024audio}) leverage deep recurrent structures to capture long-term dependencies. However, due to their sequential processing mechanism \cite{lei2018sru}, their computational efficiency is limited. Transformer-based models, represented by AV-Sepformer \cite{lin2023av}, achieve synchronous modeling of modal information through a dual-path Transformer architecture, which substantially improves separation performance. Nonetheless, the computational complexity of such models scales quadratically along with sequence length, leading to increased computational burden and model size. Lastly, hybrid architecture models (e.g., AV-TF-GridNet \cite{AV-TFGridNet} and RTFS-Net \cite{pegg2023rtfs}) integrate CNN, Transformer, and RNN modules to balance between performance and computational complexity, achieving superior separation results in offline speech separation tasks. However, these methods typically rely on global information, thus limiting their application in real-time separation scenarios. In this paper, we aim to develop a causal AVSS method capable of operating under streaming processing conditions.

\subsection{Causal Models}

In the audio-only streaming domain, several causal architectures—such as Conv-TasNet \cite{luo2019conv}, SkiM \cite{li2022skim}, and ReSepformer \cite{della2024resource}—have demonstrated impressive real-time separation performance. However, when directly applied to the AVSS domain, these methods reveal critical limitations: since they operate exclusively on acoustic features, they fail to leverage complementary visual cues; moreover, under causal constraints, their strict frame-by-frame fusion and limited recurrence hinder the modeling of long-range, cross-modal temporal dependencies, ultimately degrading robustness in visually guided scenarios.

In the AVSS domain, compared to non-causal methods, research on causal methods remains in its early stages. Some researchers \cite{zhang2021online} have even attempted to adapt the causal Conv-TasNet \cite{luo2019conv} to the AVSS task, but the performance gap compared to non-causal AVSS methods remains substantial. Therefore, these audio-only causal methods cannot be directly and effectively transferred to the AVSS task without significant redesign. Under causal constraints, achieving efficient cross-modal information fusion while fully leveraging historical temporal dependencies remains a critical challenge in causal AVSS tasks. In this paper, we propose Swift-Net, which efficiently integrates historical information by introducing power spectrogram features and the Grouped SRU network, thereby substantially improving our model’s separation performance.


\section{Causal Design Strategies}
\label{sec:causal}

To ensure that the process of audio-visual separation strictly adheres to temporal causality, that is, relying solely on current and past information while completely avoiding any leakage of future information, we have specifically redesigned and adjusted the neural network modules commonly employed in non-causal AVSS methods. In particular, we have implemented causal convolution, unidirectional recurrent networks, self-attention modules with masking mechanisms, and segmented causal average pooling, thereby ensuring that all components of the overall model comply with causality constraints. 
We have also accordingly modified AV-Sepformer \cite{lin2023av}, CTCNet \cite{li2024audio}, AV-DPRNN \cite{tao2024audio}, AV-TF-GridNet \cite{AV-TFGridNet}, and RTFSNet \cite{pegg2023rtfs}, and conducted a systematic comparison with Swift-Net. 

\textbf{Causal Convolutional Layers.} When the kernel size of a convolution exceeds 1, standard convolution operations inevitably introduce future information when computing the output for the current frame. By applying causal padding prior to the convolution operation \cite{harell2019wavenilm,ma2021ecg}, convolutional layers can be made causal.

\textbf{Causal RNN Layer.} For RNN layers like LSTM \cite{hochreiter1997long}, SRU \cite{lei2018sru} and GRU \cite{chung2014empirical}, causal processing is inherently ensured by setting the network to operate unidirectionally, guaranteeing that only past information is utilized. 

\textbf{Causal Attention Layer.} The causal attention layer is implemented by introducing an upper triangular mask into the self-attention mechanism, consistent with approaches adopted in existing causal large language models \cite{chang2024survey}. Specifically, we construct an upper triangular mask matrix where all elements above the main diagonal are assigned a value of negative infinity. This design ensures that the attention computation at each timestep depends only on the current and previous information, thereby strictly adhering to causality constraints.

\begin{figure}[h]
\centering
\includegraphics[width=0.3\textwidth]{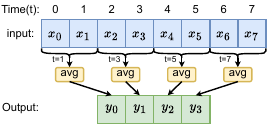}
\caption{Diagram for segment-based causal adaptive average pooling layer. For clarity of presentation, we take an input sequence of length $T=8$ that is evenly divided into $S=4$ segments as an example. Here, \texttt{avg} denotes the averaging operation. At $t=1$, the computation begins when $x_1$ arrives; at $t=3$, the computation begins when $x_3$ arrives, and so on.
}
\label{fig:causalpooling.png}
\end{figure}

\textbf{Causal Average Pooling Layer.} We propose a segment-based causal adaptive average pooling method, as shown in Figure~\ref{fig:causalpooling.png}. Specifically, let the input sequence be $\mathbf{X} \in \mathbb{R}^{1 \times T}$, where $T$ denotes the total number of timesteps. We first uniformly divide $\mathbf{X}$ along the temporal axis into $N$ non-overlapping segments, each with length $L = \lceil T/N \rceil$. For the $n$-th output interval ($n \in \{1,2,\dots,N\}$), the corresponding temporal range is $[(n-1)L+1,\, nL]$. To satisfy the causality constraint during the pooling operation, the pooling result of the $n$-th segment, $\mathbf{y}_n$, depends only on the current and previous inputs, that is,
\begin{equation}
    \mathbf{y}_n = \frac{1}{nL} \sum_{t=1}^{nL} \mathbf{x}_t.
\end{equation}
This way, the $n$-th output captures not only the information within the current segment but also the historical information from all preceding segments. Compared with conventional average pooling, this "causal pooling" strictly ensures that the output does not utilize any future information, thereby meeting the causality requirements essential for sequential modeling. 

\section{Method}

\begin{figure*}[h]
\centering
\includegraphics[width=0.8\textwidth]{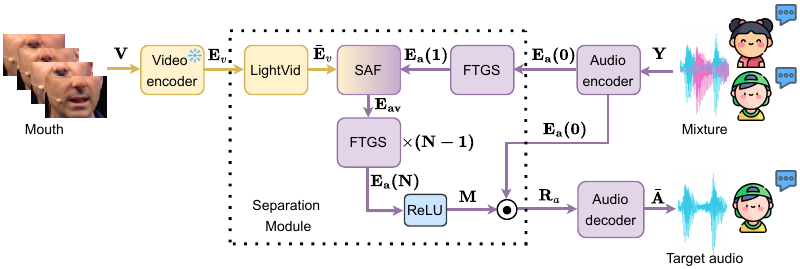}
\caption{The overall pipeline of Swift-Net. The yellow line and the purple line represent the flow of visual features and audio features respectively. The $\odot$ symbol denotes element-wise multiplication in the complex domain.}
\label{fig:All.png}
\vspace{-10pt}
\end{figure*}

Our proposed causal AVSS model, Swift-Net, consists of an audio encoder, a video encoder, a separation module, and an audio decoder, as illustrated in Figure~\ref{fig:All.png}. Specifically, the input comprises a mixed audio signal $\mathbf{Y} \in \mathbb{R}^{1 \times T_a}$ and the corresponding grayscale lip movement video sequence $\mathbf{V} \in \mathbb{R}^{1 \times T_v \times H \times W}$, where $T_a$ denotes the length of the audio sequence, and $T_v$, $H$, and $W$ represent the number, height, and width of video frames, respectively. First, the audio encoder maps the mixed audio signal to an audio embedding $\mathbf{E_a(0)} \in \mathbb{R}^{C_a \times T'_a \times F}$, and the video encoder maps the lip movement video sequence to a lip embedding $\mathbf{E}_v \in \mathbb{R}^{C_v \times T_v}$, where $C_a$ and $C_v$ denote the dimensions of the encoded features, and $T'_a$ is the corresponding encoded temporal lengths, $F$ denotes the number of frequency bins in the spectrogram. Subsequently, a causal audio-visual separation network generates an estimated mask $\mathbf{M} \in \mathbb{R}^{C_a \times T'_a \times F}$ for the target speaker based on $\mathbf{E_a(0)}$ and $\mathbf{E}_v$. Finally, following the approach of RTFSNet \cite{pegg2023rtfs}, the audio decoder performs element-wise multiplication of $\mathbf{E_a(0)}$ and $\mathbf{M}$ in the complex domain to obtain the audio embedding of the target speaker, $\mathbf{R}_a \in \mathbb{R}^{C_a \times T'_a \times F}$. $\mathbf{R}_a$ is transformed back to the time-domain waveform by applying the decoder, thus yielding the separated speech of the target speaker $\bar{\mathbf{A}} \in \mathbb{R}^{1 \times T_a}$.

\subsection{Encoder}

For the video encoder $L_m(\cdot)$, we employ the pretrained CTCNet-Lip \cite{li2024audio} model to independently extract visual features $\mathbf{E}_v$ of the target speaker from each lip movement video sequence $\mathbf{V} \in \mathbb{R}^{1 \times T_v \times H \times W}$:
\begin{equation}
\mathbf{E}_v = L_m(\mathbf{V}), \quad  
\mathbf{E}_v \in \mathbb{R}^{C_v \times T_v}.
\end{equation}

For the audio encoder, the input mixed audio signal $\mathbf{Y}$ is first processed by a Short-Time Fourier Transform (STFT) to obtain its real $\mathbf{Y}_r\in \mathbb{R}^{T'_a\times F}$ and imaginary $\mathbf{Y}_i\in \mathbb{R}^{T'_a\times F}$ spectrograms. During this process, padding is applied to prevent the window function from accessing future information. We observed that the power spectrogram $\mathbf{G}$ contains crucial information closely related to auditory perception, effectively reflecting the power distribution of the signal \cite{hsu2004modulation},
\begin{equation}
\mathbf{G} = \sqrt{\mathbf{Y}_r^2 + \mathbf{Y}_i^2},\quad
\mathbf{G} \in \mathbb{R}^{T'_a \times F},
\end{equation}
where $F$ and $T'_a$ denote the frequency and temporal dimensions, respectively. To further enrich the audio representation, $\mathbf{G}$, $\mathbf{Y}_r$, and $\mathbf{Y}_i$ are stacked along the channel dimension to construct a composite audio feature $\mathbf{Y}_m \in \mathbb{R}^{3 \times T'_a \times F} = \{\mathbf{G} || \mathbf{Y}_r || \mathbf{Y}_i\}$, where $||$ denotes concatenation along the channel dimension. Thereafter, $\mathbf{Y}_m$ is fed into a convolutional layer with normalization and nonlinear activation to obtain the audio embedding $\mathbf{{E_a}(0)}$, which is subsequently input to the following audio-visual separation network.

\subsection{The Separation Module}

The overall framework comprises three core components: a lightweight video processing block (LightVid block), an efficient audio block (FTGS block), and a selective attention audio-visual fusion block (SAF block). First, visual features $\mathbf{E}_v$ generated by the video encoder feed into the LightVid block to obtain enhanced visual embeddings $\bar{\mathbf{E}}_v$. Simultaneously, audio features $\mathbf{E_a(0)}$ are processed by the FTGS block to generate refined audio embeddings $\mathbf{E_a(1)}$. Subsequently, features from both modalities $\{\bar{\mathbf{E}}_v \in \mathbb{R}^{C_v \times T_v}, \mathbf{{E}_{a}(1)} \in \mathbb{R}^{C_a \times T'_a\times F}\}$ flow into the SAF block to produce a joint audio-visual representation $\mathbf{{E}_{av}} \in \mathbb{R}^{C_a \times T'_a\times F}$. Next, $\mathbf{\mathbf{E}_{av}}$ is fed into a deep network composed of $N-1$ stacked Efficient FTGS blocks, where each FTGS block shares parameters to further integrate and refine the joint representation, ultimately yielding iteratively optimized features $\mathbf{{E}_{a}(N)} \in \mathbb{R}^{C_a \times T'_a\times F}$.  
This cascaded structure, combined with the parameter sharing mechanism, significantly enhances the computational efficiency and effectively reduces the model size, enabling the separator to achieve real-time, low-latency inference while maintaining superior separation performance.

\subsubsection{The LightVid block}
\begin{figure}[h]
\centering
\includegraphics[width=7cm]{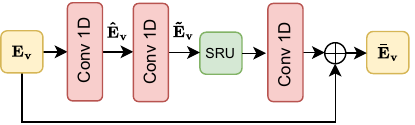}
\caption{The structural diagram of LightVid block. Here, $c_v$ denote the number of channels of the visual features, and $T_v$ represents the number of frames of the visual features.
}
\label{fig:LigthVidBlock.png}
\end{figure}

Most of the current mainstream AVSS models employ complex visual encoders involving delicate usage of deep convolutional neural networks or Transformer \cite{vaswani2017attention} architectures, often entailing considerable computational cost. To address this issue, 
we propose a lightweight video processing module, termed LightVid block (see Figure~\ref{fig:LigthVidBlock.png}), which serves as an efficient causal visual feature extractor capable of extracting both local spatial information and long-range temporal dynamics from lip-reading video sequences with low latency.

Specifically, the LightVid block combines depthwise separable convolutional layers with linear recurrent units to efficiently process input visual feature sequence. For each frame, given the lip region features $\{\mathbf{E}_{v,i} \in \mathbb{R}^{C_v}\mid i\in[1,T_v]\}$, a $1\times 1$ convolutional layer followed by layer-normalization is first applied to independently model $\mathbf{E}_{v,i}$ along the channel dimension, resulting in the processed features $\hat{\mathbf{E}}_v \in \mathbb{R}^{C_v \times T_v}$. This operation effectively preserves the local semantic information of lip movements and provides a foundation for subsequent temporal modeling. Afterwards, another $1\times 1$ convolution is used to project $\hat{\mathbf{E}}_v$ to a lower-dimensional space, yielding $\tilde{\mathbf{E}}_v \in \mathbb{R}^{C_h \times T_v}$, where $C_h < C_v$, so as to reduce the number of feature channels and improve computational efficiency. Next, a unidirectional SRU is leveraged to model the temporal dependencies and dynamic variations of lip movements across consecutive frames based on $\tilde{\mathbf{E}}_v$. Finally, a $1\times 1$ convolution projects the SRU outputs back to the original feature dimension, and a residual connection from the block input is incorporated. The final output is the enhanced visual feature representation $\bar{\mathbf{E}}_v \in \mathbb{R}^{C_v\times T_v}$.

\begin{figure*}[h]
\centering
\includegraphics[width=0.95\textwidth]{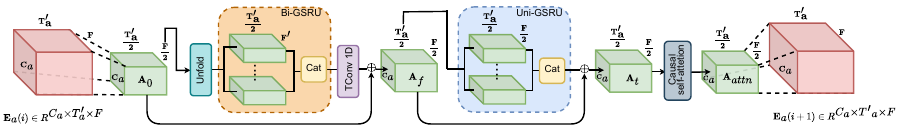
}
\caption{The structural diagram of FTGS block. Bi-GSRU denotes the Bidirectional Grouped SRU. UNi-GSRU denotes the Unidirectional Grouped SRU.
}
\label{fig:FTGS block.png}
\vspace{-10pt}
\end{figure*}

\subsubsection{The FTGS block}

Some existing time-frequency domain AVSS methods \cite{pegg2023rtfs,AV-TFGridNet} employ two RNNs to model the audio signal in the temporal and frequency dimensions, respectively. As a result, their computational complexity is approximately twice that of time-domain AVSS methods \cite{li2023iianet,li2024audio}, often leading to significant inference latency, making it challenging to meet real-time processing requirements. To address this issue, we propose an efficient Frequency-Time Grouped SRU block (FTGS block). In the FTGS block, the channels of audio features are divided into multiple groups, each of which is independently modeled by lightweight recurrent units with fewer parameters. The outputs of all groups are subsequently concatenated along the channel dimension. The grouping strategy decomposes the overall modeling task into several parallelizable sub-tasks, thereby substantially reducing computational overhead.

Figure~\ref{fig:FTGS block.png} shows a structural diagram of the FTGS block. Let the auditory feature representation input to the FTGS block be denoted as $\mathbf{E}_{a}(i) \in \mathbb{R}^{C_a \times T'_a \times F}$. First, a $1 \times 1$ two-dimensional convolution is applied to $\mathbf{E}_{a}(i)$ for downsampling, resulting in $\mathbf{A}_0 \in \mathbb{R}^{C_a \times \frac{T'_a}{2} \times \frac{F}{2}}$. Subsequently, following a strategy similar to TF-GridNet \cite{wang2023tf}, we first unfold $\mathbf{A}_0$ along the frequency dimension using a kernel size of 8 and a stride of 1 to obtain $\mathbf{A}_0' \in \mathbb{R}^{8C_a \times \frac{T'_a}{2} \times {F'}}$ , where $\mathbf F'$ is the resulting unfolded frequency dimension. It then gets divided into $G$ groups along the channel dimension, denoted as $\mathbf{A'}_0 = [\mathbf{A'}_0^{(1)}; \mathbf{A'}_0^{(2)}; \cdots; \mathbf{A'}_0^{(G)}]$, where $\mathbf{A'}_0^{(g)} \in \mathbb{R}^{C_a^* \times \frac{T'_a}{2} \times {F'}}$ and $C_a^* = 8C_a / G$ is the number of channels of each group. For each group $\mathbf{A'}_0^{(g)}$ and at each frame $t$, the features corresponding to different frequencies are treated as a sequence along the frequency axis. A bidirectional SRU is employed to capture frequency correlations as:
\begin{equation*}
    \tilde{\mathbf{X}}^{(g)}_{c, :, t} = \mathrm{BiSRU}\big(\mathbf{A'}_0^{(g)}[c, t, :]\big) \quad \forall c = 1, \ldots, C_a^*,\ t = 1, \ldots, \frac{T'_a}{2}.
\end{equation*}
This procedure enhances the model's ability to capture inter-frequency coupling relationships. The outputs of all $G$ groups are then concatenated along the channel dimension to restore the original number of channels, yielding \[\mathbf{\tilde{A}}_0 = \mathop{\mathrm{Concat}}\left(\tilde{\mathbf{X}}^{(1)}, \tilde{\mathbf{X}}^{(2)}, \dots, \tilde{\mathbf{X}}^{(G)}\right).\]
After upsampling the feature map along the frequency dimension to the original resolution via a transposed convolution, a residual connection is then applied via element-wise addition with $\mathbf{A}_0$ to obtain $\mathbf{A}_f \in \mathbb{R}^{C_a \times T'_a \times F}$.

In contrast to frequency modeling, temporal modeling employs unidirectional grouped SRUs to ensure a causal structure, enhancing the integration of information from historical feature spaces and thereby improving separation performance. The output is denoted as $\mathbf{A}_t$. Next, a causal self-attention mechanism is applied to $\mathbf{A}_t$ to further improve the audio feature representation, resulting in $\mathbf{A}_{att}$. Finally, a reconstruction strategy similar to TDANet \cite{li2022efficient} is adopted to recover the time-frequency resolution, generating the next-level feature $\mathbf{E}_{a}(i+1)$.

In our FTGS Block architecture, we adopt grouped SRU units (Grouped SRUs) based on a grouping strategy. Compared to the standard SRU \cite{lei2018sru}, Grouped SRU significantly reduces both parameter count and computational complexity. Specifically, let the input feature dimension be $D_\text{in}$ and the hidden state dimension be $D_\text{hid}$. The main parameters of a standard SRU come from the input-to-hidden transformation (e.g., a weight matrix of size $D_\text{in} \times k D_\text{hid}$) as well as the hidden-to-hidden connections (e.g., parameters of size $D_\text{hid} \times l D_\text{hid}$). Ignoring bias terms, the total parameter count can be approximated as $P_\text{SRU} \approx c_1 D_\text{in} D_\text{hid} + c_2 D_\text{hid}^2$, where $c_1$ and $c_2$ are constants related to the internal structure of the SRU. Similarly, the primary computation cost $C_\text{SRU}$, measured in MACs, is dominated by matrix multiplications and has a similar structure, $C_\text{SRU} \propto D_\text{in} D_\text{hid} + D_\text{hid}^2$.
For Grouped SRU, we partition both the input and hidden dimensions into $G$ groups, with the subspace dimensions of each group being $D_\text{in}/G$ and $D_\text{hid}/G$, respectively, and run a separate sub-SRU independently in each group. The parameter count for a single sub-SRU is $c_1 (D_\text{in}/G)(D_\text{hid}/G) + c_2 (D_\text{hid}/G)^2$, yielding a total parameter count of $P_\text{GSRU} = G \times P_{sub\text{-}SRU} \approx \frac{c_1 D_\text{in} D_\text{hid}}{G} + \frac{c_2 D_\text{hid}^2}{G}$. Thus, both the total parameter count and computation cost are reduced approximately by a factor of $1/G$:
\begin{equation}
    \frac{P_\text{GSRU}}{P_\text{SRU}} \approx \frac{C_\text{GSRU}}{C_\text{SRU}} \approx \frac{1}{G}.
\end{equation}
This result clearly indicates that, as the group number $G$ increases, the storage and computational costs of the model decrease inversely.

\subsubsection{The SAF block}
\begin{figure}[hbtp]
\centering
\includegraphics[width=7cm]{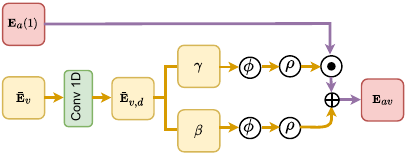}
\caption{The structural diagram of SAF block. The yellow line and the purple line represent the flow of visual features and audio features respectively. $\phi$ represents nearest neighbor interpolation. $\rho$ represents the flattening of the last dimension of the tensor. }
\label{fig:SAFBlock}
\end{figure}

Effective fusion of audio-visual features is crucial for AVSS performance, yet existing methods typically rely on complex fusion strategies that are computationally intensive. To address this, we propose a streamlined SAF block (see Figure \ref{fig:SAFBlock}) that adaptively integrates visual cues into audio features with high efficiency.
Specifically, inspired by the IIANet \cite{li2023iianet}, the SAF block implements a channel-wise selective attention strategy. First, the visual features $\bar{\mathbf{E}}_v$ extracted by the LightVid block are passed through a $1\times1$ convolution, yielding features of shape $\bar{\mathbf{E}}_{v,d} \in \mathbb{R}^{2C_v \times T_v}$. Subsequently, $\bar{\mathbf{E}}_{v,d}$ is equally divided along the channel dimension into $\gamma \in \mathbb{R}^{C_v \times T_v}$ (scaling factor) and $\beta \in \mathbb{R}^{C_v \times T_v}$ (bias term), which are used for subsequent scaling and shifting operations on the audio features, respectively. Considering potential differences in temporal scale or frame rate between visual and audio features, nearest-neighbor interpolation is utilized to align $\gamma$ and $\beta$ along the temporal axis to match the time steps of the audio features, $T'_a$. Following this alignment, the audio features $\mathbf{E}_a(1) \in \mathbb{R}^{C_a \times T'_a \times F}$ are modulated in a channel-wise manner as follows:
\begin{equation}
    \mathbf{E}_{av} = \mathbf{E}_{a}(1) \odot \gamma + \beta,
\end{equation}
where $\odot$ denotes element-wise multiplication. The resulting fused feature map $\mathbf{E}_{av} \in \mathbb{R}^{C_a \times T'_a \times F}$ incorporates both audio and visual information at each time step, thereby enabling a more effective emphasis on the target speaker's speech representations. 

\subsection{Masking and Decoder}

Consistent with RTFSNet \cite{pegg2023rtfs}, we adopt a complex-domain masking strategy for audio feature reconstruction, modified to operate causally. Specifically, given the refined feature $\mathbf{E}_{a}(N) \in \mathbb{R}^{C_a \times T \times F}$, we generate a mask $\mathbf{M} \in \mathbb{R}^{C_a \times T \times F}$ using a convolutional module with PReLU nonlinearity. Subsequently, the audio mixture embedding $\mathbf{E}_{a} \in \mathbb{R}^{C_a \times T \times F}$ and the generated mask $\mathbf{M}$ are each decomposed into their real and imaginary parts, i.e., $\mathbf{E}_a(0) = \mathbf{E}_r + j\mathbf{E}_i$ and $\mathbf{M} = \mathbf{M}_r + j\mathbf{M}_i$, where $\mathbf{E}_r, \mathbf{E}_i, \mathbf{M}_r, \mathbf{M}_i \in \mathbb{R}^{C_a \times T \times F}$. Next, we perform feature fusion using the rules of complex multiplication: $\mathbf{R}_r = \mathbf{E}_r \odot \mathbf{M}_r - \mathbf{E}_i \odot \mathbf{M}_i$ and $\mathbf{R}_i = \mathbf{E}_r \odot \mathbf{M}_i + \mathbf{E}_i \odot \mathbf{M}_r$, where $\odot$ denotes element-wise (Hadamard) multiplication, resulting in the separated feature $\mathbf{R}_a = \mathbf{R}_r + j\mathbf{R}_i$. Operations in the complex domain allow richer preservation of phase information in speech signals, thereby improving separation performance.

After obtaining the separated feature $\mathbf{R}_a \in \mathbb{C}^{C_a \times T \times F}$, we employ a single-layer causal transposed two-dimensional convolution to map it to a two-channel real-valued tensor, where the output channels are set to $2$, corresponding to the real and imaginary parts respectively. Finally, the output of the convolution is restored to the time domain by inverse short-time Fourier transform (iSTFT), yielding the separated waveform $\bar{\mathbf{A}} \in \mathbb{R}^{1 \times T_a}$, thereby achieving high-quality reconstruction of the target speaker's speech.


\begin{table*}[ht]
    \caption{Separation quality and computational efficiency comparison of different causal models across multiple datasets. All models are implemented with causal design strategies to ensure causality.}
    \centering
    \fontsize{8.5}{8.5}
    \selectfont
    \begin{tabular}{c|cc|cc|cc|ccc}
    \toprule
    \multirow{2}{*}{Model} 
      & \multicolumn{2}{c|}{LRS2-2Mix} 
      & \multicolumn{2}{c|}{LRS3-2Mix} 
      & \multicolumn{2}{c|}{VoxCeleb2-2Mix} 
      & Params & MACs & Time \\
      & SI-SNRi & SDRi & SI-SNRi & SDRi & SI-SNRi & SDRi & (M) & (G) & (ms) \\ 
    \midrule
    AV-ConvTasNet   & 9.8  & 10.3 & 12.5 & 12.8 & 6.9  & 8.2  & 11.0 & 15.0  & 16.1 \\
    AV-Sepformer    & 10.6 & 11.1 & 13.4 & 13.7 & 6.9  & 8.3  & 39.7 & 226.5 & 60.0 \\
    CTCNet          & 8.0  & 8.8  & 9.4  & 10.2 & 2.5  & 4.7  & 6.8  & 150.1 & 162.8 \\
    AV-DPRNN        & 10.2 & 10.7 & 11.9 & 12.3 & 7.8  & 9.0  & 1.3  & 4.5   & 5.8  \\
    AV-TF-GridNet   & 12.7 & 13.1 & 14.3 & 14.8 & 5.8  & 7.7  & 6.2  & 213.8 & 567.8 \\
    RTFSNet         & 12.4 & 12.8 & 14.0 & 14.4 & 8.4  & 10.7 & 0.6  & 49.8  & 178.8 \\ 
    \midrule
    Swift-Net-6     & 13.4 & 13.6 & 14.8 & 15.1 & 10.8 & 11.9 & \textbf{0.5} & 20.6  & 89.7  \\
    Swift-Net-9     & 13.6 & 13.8 & 15.4 & 15.8 & 11.4 & 12.4 & \textbf{0.5} & 28.6  & 130.7 \\
    Swift-Net-12    & \textbf{13.9} & \textbf{14.1} & \textbf{16.0} & \textbf{16.3} & \textbf{11.8} & \textbf{12.6} & \textbf{0.5} & 36.6  & 172.1 \\
    \bottomrule
    \end{tabular}
    \label{tab:results_full}
\end{table*}

\section{Experiments}

\subsection{Datasets}

In our experiments, we employed three publicly available audio-visual mixture benchmark datasets: LRS2-2Mix \cite{afouras2018deep}, LRS3-2Mix \cite{afouras2018lrs3}, and VoxCeleb2-2Mix \cite{chung2018voxceleb2}. To ensure consistency with existing research, we followed the same data pre-processing pipeline \cite{li2024audio,pegg2023rtfs}. For video processing, all video samples were normalized to 25 frames per second (FPS). Subsequently, utilizing a face detection network \cite{zhang2016joint}, we extracted the lip region of the target speaker from each frame and uniformly cropped it to grayscale images of size $96 \times 96$. Regarding audio processing, the sampling rate of all speech samples was set to 16 kHz. To synchronize with the video data, we extracted 2-second mixed audio segments from each sample, corresponding to 50 frames of video.

\subsection{Implementation Details}

In the FTGS block, we set different numbers of repeated iterations $N \in \{6, 9, 12\}$, and denoted the corresponding models as Swift-Net-\{6, 9, 12\}. Except for the number of repetitions, all other hyperparameters were kept consistent across configurations. In the audio encoder of Swift-Net, we utilized STFT/iSTFT with window length 256 under Hann window function, and hop size 128, to segment the audio signal and obtain a 129-dimensional spectral representation for each frame. For the LightVid block, the hidden size of the SRU was set to 64. Within the FTGS block, downsampling and upsampling operations were performed once along both the temporal and frequency dimensions at a factor of 2. In the $F$ domain, the FTGS block employed a bidirectional Grouped SRU with a hidden size of 32, whereas in the $T$ domain, a unidirectional Grouped SRU with a hidden size of 64 was used. Moreover, the number of heads in the causal self-attention mechanism was set to 4.

For training, we employed an early stopping strategy with a maximum number of 500 epochs. All experiments were conducted on 8 NVIDIA H800 GPUs with batch size 8. The optimizer used was AdamW \cite{loshchilov2017decoupled} with weight decay $1 \times 10^{-1}$ and initial learning rate $1 \times 10^{-3}$. If the validation loss did not reach a new low for 5 consecutive epochs, the learning rate was halved. The loss function was defined as the scale-invariant signal-to-noise ratio (SI-SNR) \cite{le2019sdr} between the predicted and target speech signals.

To comprehensively evaluate model performance, we assessed both separation quality and computational efficiency. The separation quality was measured using signal-to-distortion ratio improvement (SDRi) \cite{vincent2006performance} and SI-SNR improvement (SI-SNRi) \cite{le2019sdr}. For model complexity and computational efficiency, we reported the number of parameters (Params, in millions), the number of multiply-accumulate operations (MACs, in billions), and end-to-end inference time (in milliseconds). All computational cost metrics were evaluated on an NVIDIA 2080 GPU, using 2-second audio segments for speech separation, to ensure fair comparisons across different methods.


\subsection{Comparison with Existing Methods}
\label{sec:results}

We conducted a systematic evaluation of Swift-Net on three datasets and compared it with various mainstream AVSS models. Results are shown in Table~\ref{tab:results_full}. To ensure fairness, all comparison methods were adapted to their causal versions following the causal strategy described in Section \ref{sec:causal}, and the open-source codes of each model were available\footnote{\url{https://github.com/JusperLee/Swift-Net}}. Moreover, all compared models were trained within the same training framework and evaluated using consistent performance metrics. Across all datasets, it was noteworthy that Swift-Net-6, as the fastest inference variant, achieved an SI-SNRi of 13.4 dB, outperforming existing SOTA causal methods such as causal AV-TF-GridNet. In addition, Swift-Net-6 reduced the parameter count by 11.8 times and the computational cost by 10.3 times compared with causal AV-TF-GridNet. Furthermore, our high-performance variant, Swift-Net-12, achieved over 1 dB improvement in performance relative to causal AV-TF-GridNet. These results collectively demonstrated that Swift-Net offered significant advantages for causal speech separation tasks.

In addition, we further conducted a comparative analysis of the separation performance of RTFSNet \cite{pegg2023rtfs}, AV-TF-GridNet \cite{AV-TFGridNet}, and Swift-Net in real-world multi-speaker scenarios. Specifically, we collected four distinct multi-speaker video samples from the YouTube platform. To facilitate intuitive comparison of the separated speech results, we developed an interactive demo website\footnote{\url{https://cslikai.cn/Swift-Net/}} allowing users to listen and compare the outputs. Subjective listening evaluations indicated that, compared to RTFSNet and AV-TF-GridNet, Swift-Net produces speech signals that were clearer, more natural, effectively suppressing background interference.

\subsection{Ablation Studies}

We conducted ablation studies on the LRS2-2Mix dataset based on Swift-Net-6 to analyze the effectiveness of key modules and design strategies in Swift-Net, thereby verifying the rationality of the proposed method.

\textbf{Different Visual Feature Extraction Modules}.
We replaced the LightVid block visual‐feature encoder with those from ConvTasNet and RTFSNet while keeping every other component of the streaming AVSS pipeline identical in order to quantify its module‐level benefits. As shown in Table~\ref{tab:videoblock}, this lean design requires only 67.27 K parameters and 3.39 M MACs -- over 4× reduction relative to ConvTasNet and RTFSNet. Remarkably, despite its minimal complexity, LightVid block still delivers a slight performance edge, confirming that a structurally simple encoder can unite efficiency and effectiveness in real-time causal AVSS.

\begin{table}[htbp]
    \centering
    \caption{Results of using different visual feature extraction modules. The parameters and MACs represent those used only during the  visual feature extraction process.}
    \fontsize{8.5}{8.5}
    \selectfont
    \begin{tabular}{c|c|c|c|c}
    \toprule
    Video block & SI-SNRi & SDRi & Params (K) & MACs (M)  \\ 
    \midrule
    ConvTasNet & 13.05 & 13.37 & 265.73 & 13.34 \\
    RTFSNet & 13.15 & 13.45 & 284.55 & 7.63 \\
    \midrule
    \makecell{LightVid block \\ (ours)} & \textbf{13.34} & \textbf{13.64} & \textbf{67.27} & \textbf{3.39} \\
    \bottomrule
    \end{tabular}
    \label{tab:videoblock}
\end{table}

\textbf{Number of the Grouped SRUs}. To investigate the effect of the number of Grouped SRUs on the model, we tested using \{1, 2, 4, 8\} Grouped SRUs within the FTGS block. The experimental results in Table~\ref{tab:results_group} indicated that, in terms of separation performance, using two Grouped SRUs achieved similar performance to the ungrouped setting (i.e., group number being one). However, in terms of model efficiency, models with Grouped SRUs demonstrated significant advantages: they reduced the number of parameters by approximately 15\%, and decreased the computational cost by approximately 19\%. In addition, grouping enabled parallel processing of channels along both time and frequency dimensions, enabling more effective utilization of historical information and more efficient capture of long-term temporal dependencies. This mechanism substantially improved overall model efficiency, which is crucial for real-time applications.

\begin{table}[ht]
    \caption{Results of using different group sizes. Parameters and MACs indicate the entire network structure.}
    \centering
    \fontsize{8.5}{8.5}
    \selectfont
    \begin{tabular}{c|c|c|c|c}
    \toprule
    Number & SI-SNRi & SDRi & Params (M) & MACs (G)  \\ 
    \midrule
    1     & \textbf{13.56} & \textbf{13.82} & 0.63 & 25.57 \\
    2 & 13.34 & 13.64 & 0.53 & 20.68 \\
    4 & 12.85 & 13.16 & 0.47 & 18.23 \\
    8 & 12.51 & 12.86 & \textbf{0.44} & \textbf{17.01} \\
    \bottomrule
    \end{tabular}
    \vspace{4pt}
    \label{tab:results_group}
\end{table}

\textbf{Comparison of Different RNN Types in Group Modules.} We conducted a comparative analysis on how different recurrent units in the group module affect model performance. In all experiments, the number of groups was set to 2. The default SRU was respectively replaced with GRU \cite{cho2014properties}, LSTM \cite{hochreiter1997long}, and vanilla RNN, and the results were reported in Table \ref{tab:results_rnn}. Experimental results showed that the SRU unit achieved the best performance on the separation task. Other more complex RNN structures such as GRU and LSTM led to a decrease in SI-SNRi of more than 0.4 dB. Furthermore, the model based on SRU maintained a parameter count of only 0.53M, indicating an optimal balance between performance and efficiency. Thus, adopting SRU in the group module effectively enhanced separation performance while maintaining low computational and storage costs.

\begin{table}[ht]
    \caption{Results of using different types of RNNs based on grouping.}
    \centering
    \fontsize{8.5}{8.5}
    \selectfont
    \begin{tabular}{c|c|c|c|c}
    \toprule
    RNN Type & SI-SNRi & SDRi & Params (M) & MACs (G) \\ 
    \midrule
    GRU & 13.21 & 13.50 & 0.55 & 22.02 \\
    LSTM & 12.92 & 13.24 & 0.59 & 24.11 \\
    RNN  & 12.12 & 12.50 & 0.44 & 17.83 \\
    SRU (ours) & \textbf{13.34} & \textbf{13.64} & \textbf{0.53} & \textbf{20.68} \\
    \bottomrule
    \end{tabular}
    \label{tab:results_rnn}
\end{table}

\textbf{Comparison of Different Fusion Module Designs}. We conducted a systematic comparison of various design strategies for the audio-visual fusion module to evaluate their specific impact on separation performance. In Swift-Net, the default strategy adopted a selective attention-based fusion mechanism. For comprehensive analysis, we compared this approach with the following three alternatives: (1) Concatenation Fusion: the aligned visual features were concatenated with audio features along the channel dimension, followed by convolutional layers for fusion; (2) Addition Fusion: visual and audio features were directly added element-wise; (3) Cross-Attention Fusion: the fusion was implemented through a cross-attention mechanism.

As shown in Table~\ref{tab:results_fusion}, the selective attention-based fusion achieved the best separation performance, indicating that other schemes failed to fully utilize the visual information. It was noteworthy that although the cross-attention fusion theoretically possessed stronger modeling capability, its actual performance was significantly inferior to ours. Moreover, its parameter count exceeded 26.38M, and computational complexity reached 27.24G MACs, which was not suitable for real-time applications. By contrast, our method maintained almost the same parameter size and computational complexity as the direct addition and cascaded convolutional fusion methods; specifically, the parameter count was constrained to approximately 0.53M, and the computation was below 0.1G MACs. 

\begin{table}[htbp]
    \centering
    \caption{Results of using different fusion strategies.}
    \fontsize{8.5}{8.5}
    \selectfont
    \begin{tabular}{c|c|c|c|c}
    \toprule
    Fusion Strategy & SI-SNRi & SDRi & Params (M) & MACs (G)  \\ 
    \midrule
    Concatenation & 12.85 & 13.15 & 0.72 & 27.09 \\
    Addition & 12.73 & 12.97 & 0.64 & 20.69 \\
    Cross-attention & 13.10 & 13.40 & 26.38 & 27.24 \\
    SAF (ours) & \textbf{13.34} & \textbf{13.64} & \textbf{0.53} & \textbf{20.68} \\
    \bottomrule
    \end{tabular}
    \label{tab:results_fusion}
\end{table}

\textbf{Effect of Incorporating Power Spectrogram Features.} We systematically evaluated the impact of introducing audio power spectrogram features on audio separation performance across three models: RTFSNet, AV-TF-GridNet, and Swift-Net. As shown in Table \ref{tab:results_power}, in Swift-Net (with 0.53M parameters and an increase of less than 0.1G MACs), the inclusion of power spectrogram features raised the SI-SNRi from 13.1 to 13.3. Furthermore, introducing power spectrogram features resulted in negligible increases in model parameters and incurred minimal MACs overhead, making it a low-cost enhancement. We conjectured that power spectrogram features provided key speech-presence cues, thereby improving audio-visual fusion and separation performance.

\begin{table}[h]
    \caption{Results with (upper) and without (lower) power spectrogram modeling on the LRS2-2Mix dataset. }
    \centering
    \fontsize{8.5}{8.5}\selectfont
    \begin{tabular}{c|c|c|c|c}
    \toprule
    Method & SI-SNRi & SDRi & Params~(M) & MACs~(G)  \\ 
    \midrule
    \makecell[cc]{RTFSNet}        & \makecell[c]{12.8\\12.4}        & \makecell[c]{13.1\\12.8}       & \makecell[c]{0.64\\0.63}       & \makecell[c]{49.89\\49.81}    \\
    \midrule
    \makecell[cc]{AV-TF-GridNet}  & \makecell[c]{13.0\\12.7}        & \makecell[c]{13.4\\13.1}       & \makecell[c]{6.26\\6.25}       & \makecell[c]{213.96\\213.88}  \\
    \midrule
    \makecell[cc]{Swift-Net-6}       & \makecell[c]{\textbf{13.3}\\13.1}& \makecell[c]{\textbf{13.6}\\13.4}& \makecell[c]{{0.53}\\0.52}& \makecell[c]{{20.68}\\20.60}\\
    \bottomrule
    \end{tabular}
    \label{tab:results_power}
\end{table}

\section{Conclusions}

This work proposes a novel causal AVSS model, Swift-Net, to address the limitations of non-causal models in real-time scenarios. To overcome this challenge, our model incorporates three key innovations. First, we design a lightweight visual feature extraction module combining convolution with SRU to effectively capture temporal dependencies from historical visual contexts. Second, we propose a lightweight audio-visual fusion module utilizing a selective attention mechanism to efficiently integrate auditory features correlated with visual cues. Third, we develop an efficient Grouped SRU module that aggregates historical information across multiple feature subspaces through a grouping strategy, significantly enhancing model efficiency. Swift-Net achieved the best separation performance on the public datasets LRS2, LRS3, and VoxCeleb2, with fewer parameters and lower computational cost.



\begin{ack}
This work was supported in part by the National Key Research and Development Program of China (No. 2021ZD0200301), the Science and Technology Project of Qinghai Province (No. 2023-QY-208), and the National Natural Science Foundation of China (No. U2341228).
\end{ack}



\bibliography{mybibfile}

\begin{thebibliography}{40}
\providecommand{\natexlab}[1]{#1}
\providecommand{\url}[1]{\texttt{#1}}
\expandafter\ifx\csname urlstyle\endcsname\relax
  \providecommand{\doi}[1]{doi: #1}\else
  \providecommand{\doi}{doi: \begingroup \urlstyle{rm}\Url}\fi

\bibitem[Afouras et~al.(2018{\natexlab{a}})Afouras, Chung, Senior, Vinyals, and Zisserman]{afouras2018deep}
T.~Afouras, J.~S. Chung, A.~Senior, O.~Vinyals, and A.~Zisserman.
\newblock Deep audio-visual speech recognition.
\newblock \emph{IEEE transactions on pattern analysis and machine intelligence}, 44\penalty0 (12):\penalty0 8717--8727, 2018{\natexlab{a}}.

\bibitem[Afouras et~al.(2018{\natexlab{b}})Afouras, Chung, and Zisserman]{afouras2018lrs3}
T.~Afouras, J.~S. Chung, and A.~Zisserman.
\newblock Lrs3-ted: a large-scale dataset for visual speech recognition.
\newblock \emph{arXiv preprint arXiv:1809.00496}, 2018{\natexlab{b}}.

\bibitem[Agrawal et~al.(2023)Agrawal, Gupta, and Garg]{agrawal2023monaural}
J.~Agrawal, M.~Gupta, and H.~Garg.
\newblock Monaural speech separation using wt-conv-tasnet for hearing aids.
\newblock \emph{International Journal of Speech Technology}, 26\penalty0 (3):\penalty0 707--720, 2023.

\bibitem[Chang et~al.(2024)Chang, Wang, Wang, Wu, Yang, Zhu, Chen, Yi, Wang, Wang, et~al.]{chang2024survey}
Y.~Chang, X.~Wang, J.~Wang, Y.~Wu, L.~Yang, K.~Zhu, H.~Chen, X.~Yi, C.~Wang, Y.~Wang, et~al.
\newblock A survey on evaluation of large language models.
\newblock \emph{ACM transactions on intelligent systems and technology}, 15\penalty0 (3):\penalty0 1--45, 2024.

\bibitem[Cho et~al.(2014)Cho, Van~Merri{\"e}nboer, Bahdanau, and Bengio]{cho2014properties}
K.~Cho, B.~Van~Merri{\"e}nboer, D.~Bahdanau, and Y.~Bengio.
\newblock On the properties of neural machine translation: Encoder-decoder approaches.
\newblock \emph{arXiv preprint arXiv:1409.1259}, 2014.

\bibitem[Chung et~al.(2014)Chung, Gulcehre, Cho, and Bengio]{chung2014empirical}
J.~Chung, C.~Gulcehre, K.~Cho, and Y.~Bengio.
\newblock Empirical evaluation of gated recurrent neural networks on sequence modeling.
\newblock \emph{arXiv preprint arXiv:1412.3555}, 2014.

\bibitem[Chung et~al.(2018)Chung, Nagrani, and Zisserman]{chung2018voxceleb2}
J.~S. Chung, A.~Nagrani, and A.~Zisserman.
\newblock Voxceleb2: Deep speaker recognition.
\newblock \emph{arXiv preprint arXiv:1806.05622}, 2018.

\bibitem[Delcroix et~al.(2018)Delcroix, Zmolikova, Kinoshita, Ogawa, and Nakatani]{delcroix2018single}
M.~Delcroix, K.~Zmolikova, K.~Kinoshita, A.~Ogawa, and T.~Nakatani.
\newblock Single channel target speaker extraction and recognition with speaker beam.
\newblock In \emph{2018 IEEE international conference on acoustics, speech and signal processing (ICASSP)}, pages 5554--5558. IEEE, 2018.

\bibitem[Della~Libera et~al.(2024)Della~Libera, Subakan, Ravanelli, Cornell, Lepoutre, and Grondin]{della2024resource}
L.~Della~Libera, C.~Subakan, M.~Ravanelli, S.~Cornell, F.~Lepoutre, and F.~Grondin.
\newblock Resource-efficient separation transformer.
\newblock In \emph{ICASSP 2024-2024 IEEE International Conference on Acoustics, Speech and Signal Processing (ICASSP)}, pages 761--765. IEEE, 2024.

\bibitem[Ephrat et~al.(2018)Ephrat, Mosseri, Lang, Dekel, Wilson, Hassidim, Freeman, and Rubinstein]{ephrat2018looking}
A.~Ephrat, I.~Mosseri, O.~Lang, T.~Dekel, K.~Wilson, A.~Hassidim, W.~T. Freeman, and M.~Rubinstein.
\newblock Looking to listen at the cocktail party: A speaker-independent audio-visual model for speech separation.
\newblock \emph{arXiv preprint arXiv:1804.03619}, 2018.

\bibitem[Gao and Grauman(2021)]{gao2021visualvoice}
R.~Gao and K.~Grauman.
\newblock Visualvoice: Audio-visual speech separation with cross-modal consistency.
\newblock In \emph{2021 IEEE/CVF Conference on Computer Vision and Pattern Recognition (CVPR)}, pages 15490--15500. IEEE, 2021.

\bibitem[Harell et~al.(2019)Harell, Makonin, and Baji{\'c}]{harell2019wavenilm}
A.~Harell, S.~Makonin, and I.~V. Baji{\'c}.
\newblock Wavenilm: A causal neural network for power disaggregation from the complex power signal.
\newblock In \emph{ICASSP 2019-2019 IEEE International Conference on Acoustics, Speech and Signal Processing (ICASSP)}, pages 8335--8339. IEEE, 2019.

\bibitem[Hochreiter and Schmidhuber(1997)]{hochreiter1997long}
S.~Hochreiter and J.~Schmidhuber.
\newblock Long short-term memory.
\newblock \emph{Neural computation}, 9\penalty0 (8):\penalty0 1735--1780, 1997.

\bibitem[Hsu et~al.(2004)Hsu, Woolley, Fremouw, and Theunissen]{hsu2004modulation}
A.~Hsu, S.~M. Woolley, T.~E. Fremouw, and F.~E. Theunissen.
\newblock Modulation power and phase spectrum of natural sounds enhance neural encoding performed by single auditory neurons.
\newblock \emph{Journal of Neuroscience}, 24\penalty0 (41):\penalty0 9201--9211, 2004.

\bibitem[Ke et~al.(2023)Ke, Wang, Hu, and Wang]{ke2023single}
S.~Ke, Z.~Wang, R.~Hu, and X.~Wang.
\newblock Single-channel multi-speakers speech separation based on isolated speech segments.
\newblock \emph{Neural Processing Letters}, 55\penalty0 (1):\penalty0 385--400, 2023.

\bibitem[Krizhevsky et~al.(2012)Krizhevsky, Sutskever, and Hinton]{krizhevsky2012imagenet}
A.~Krizhevsky, I.~Sutskever, and G.~E. Hinton.
\newblock Imagenet classification with deep convolutional neural networks.
\newblock In \emph{Advances in neural information processing systems}, volume~25, 2012.

\bibitem[Le~Roux et~al.(2019)Le~Roux, Wisdom, Erdogan, and Hershey]{le2019sdr}
J.~Le~Roux, S.~Wisdom, H.~Erdogan, and J.~R. Hershey.
\newblock Sdr--half-baked or well done?
\newblock In \emph{ICASSP 2019-2019 IEEE International Conference on Acoustics, Speech and Signal Processing (ICASSP)}, pages 626--630. IEEE, 2019.

\bibitem[Lei et~al.(2018)Lei, Zhang, Wang, Dai, and Artzi]{lei2018sru}
T.~Lei, Y.~Zhang, S.~I. Wang, H.~Dai, and Y.~Artzi.
\newblock Simple recurrent units for highly parallelizable recurrence.
\newblock In \emph{Empirical Methods in Natural Language Processing (EMNLP)}, 2018.

\bibitem[Li et~al.(2022{\natexlab{a}})Li, Yang, Wang, and Qian]{li2022skim}
C.~Li, L.~Yang, W.~Wang, and Y.~Qian.
\newblock Skim: Skipping memory lstm for low-latency real-time continuous speech separation.
\newblock In \emph{ICASSP 2022-2022 IEEE International Conference on Acoustics, Speech and Signal Processing (ICASSP)}, pages 681--685. IEEE, 2022{\natexlab{a}}.

\bibitem[Li et~al.(2022{\natexlab{b}})Li, Yang, and Hu]{li2022efficient}
K.~Li, R.~Yang, and X.~Hu.
\newblock An efficient encoder-decoder architecture with top-down attention for speech separation.
\newblock In \emph{The Eleventh International Conference on Learning Representations}, 2022{\natexlab{b}}.

\bibitem[Li et~al.(2023)Li, Yang, Sun, and Hu]{li2023iianet}
K.~Li, R.~Yang, F.~Sun, and X.~Hu.
\newblock Iianet: An intra-and inter-modality attention network for audio-visual speech separation.
\newblock \emph{arXiv preprint arXiv:2308.08143}, 2023.

\bibitem[Li et~al.(2024)Li, Xie, Chen, Yuan, and Hu]{li2024audio}
K.~Li, F.~Xie, H.~Chen, K.~Yuan, and X.~Hu.
\newblock An audio-visual speech separation model inspired by cortico-thalamo-cortical circuits.
\newblock \emph{IEEE Transactions on Pattern Analysis and Machine Intelligence}, 2024.

\bibitem[Lin et~al.(2023)Lin, Cai, Dinkel, Chen, Yan, Wang, Zhang, Wu, Wang, and Meng]{lin2023av}
J.~Lin, X.~Cai, H.~Dinkel, J.~Chen, Z.~Yan, Y.~Wang, J.~Zhang, Z.~Wu, Y.~Wang, and H.~Meng.
\newblock Av-sepformer: Cross-attention sepformer for audio-visual target speaker extraction.
\newblock In \emph{ICASSP 2023-2023 IEEE International Conference on Acoustics, Speech and Signal Processing (ICASSP)}, pages 1--5. IEEE, 2023.

\bibitem[Liu and Wang(2020)]{liu2020causal}
Y.~Liu and D.~Wang.
\newblock Causal deep casa for monaural talker-independent speaker separation.
\newblock \emph{IEEE/ACM transactions on audio, speech, and language processing}, 28:\penalty0 2109--2118, 2020.

\bibitem[Loshchilov and Hutter(2017)]{loshchilov2017decoupled}
I.~Loshchilov and F.~Hutter.
\newblock Decoupled weight decay regularization.
\newblock \emph{arXiv preprint arXiv:1711.05101}, 2017.

\bibitem[Luo and Mesgarani(2019)]{luo2019conv}
Y.~Luo and N.~Mesgarani.
\newblock Conv-tasnet: Surpassing ideal time--frequency magnitude masking for speech separation.
\newblock \emph{IEEE/ACM transactions on audio, speech, and language processing}, 27\penalty0 (8):\penalty0 1256--1266, 2019.

\bibitem[Ma et~al.(2021)Ma, Chen, Zhu, Yuan, Chen, and Shu]{ma2021ecg}
H.~Ma, C.~Chen, Q.~Zhu, H.~Yuan, L.~Chen, and M.~Shu.
\newblock An ecg signal classification method based on dilated causal convolution.
\newblock \emph{Computational and Mathematical Methods in Medicine}, 2021\penalty0 (1):\penalty0 6627939, 2021.

\bibitem[Mesgarani and Chang(2012)]{mesgarani2012selective}
N.~Mesgarani and E.~F. Chang.
\newblock Selective cortical representation of attended speaker in multi-talker speech perception.
\newblock \emph{Nature}, 485\penalty0 (7397):\penalty0 233--236, 2012.

\bibitem[Michelsanti et~al.(2021)Michelsanti, Tan, Zhang, Xu, Yu, Yu, and Jensen]{michelsanti2021overview}
D.~Michelsanti, Z.-H. Tan, S.-X. Zhang, Y.~Xu, M.~Yu, D.~Yu, and J.~Jensen.
\newblock An overview of deep-learning-based audio-visual speech enhancement and separation.
\newblock \emph{IEEE/ACM Transactions on Audio, Speech, and Language Processing}, 29:\penalty0 1368--1396, 2021.

\bibitem[Pan et~al.(2023)Pan, Wichern, Masuyama, Germain, Khurana, Hori, and Le~Roux]{AV-TFGridNet}
Z.~Pan, G.~Wichern, Y.~Masuyama, F.~G. Germain, S.~Khurana, C.~Hori, and J.~Le~Roux.
\newblock Scenario-aware audio-visual tf-gridnet for target speech extraction.
\newblock In \emph{2023 IEEE Automatic Speech Recognition and Understanding Workshop (ASRU)}, pages 1--8. IEEE, 2023.

\bibitem[Pegg et~al.(2023)Pegg, Li, and Hu]{pegg2023rtfs}
S.~Pegg, K.~Li, and X.~Hu.
\newblock Rtfs-net: Recurrent time-frequency modelling for efficient audio-visual speech separation.
\newblock \emph{arXiv preprint arXiv:2309.17189}, 2023.

\bibitem[Rahimi et~al.(2022)Rahimi, Afouras, and Zisserman]{rahimi2022reading}
A.~Rahimi, T.~Afouras, and A.~Zisserman.
\newblock Reading to listen at the cocktail party: Multi-modal speech separation.
\newblock In \emph{Proceedings of the IEEE/CVF Conference on Computer Vision and Pattern Recognition}, pages 10493--10502, 2022.

\bibitem[Tao et~al.(2024)Tao, Qian, Jiang, Li, Wang, and Li]{tao2024audio}
R.~Tao, X.~Qian, Y.~Jiang, J.~Li, J.~Wang, and H.~Li.
\newblock Audio-visual target speaker extraction with reverse selective auditory attention.
\newblock \emph{arXiv preprint arXiv:2404.18501}, 2024.

\bibitem[Vaswani et~al.(2017)Vaswani, Shazeer, Parmar, Uszkoreit, Jones, Gomez, Kaiser, and Polosukhin]{vaswani2017attention}
A.~Vaswani, N.~Shazeer, N.~Parmar, J.~Uszkoreit, L.~Jones, A.~N. Gomez, {\L}.~Kaiser, and I.~Polosukhin.
\newblock Attention is all you need.
\newblock \emph{Advances in neural information processing systems}, 30, 2017.

\bibitem[Vincent et~al.(2006)Vincent, Gribonval, and F{\'e}votte]{vincent2006performance}
E.~Vincent, R.~Gribonval, and C.~F{\'e}votte.
\newblock Performance measurement in blind audio source separation.
\newblock \emph{IEEE transactions on audio, speech, and language processing}, 14\penalty0 (4):\penalty0 1462--1469, 2006.

\bibitem[Wang et~al.(2018)Wang, Muckenhirn, Wilson, Sridhar, Wu, Hershey, Saurous, Weiss, Jia, and Moreno]{wang2018voicefilter}
Q.~Wang, H.~Muckenhirn, K.~Wilson, P.~Sridhar, Z.~Wu, J.~Hershey, R.~A. Saurous, R.~J. Weiss, Y.~Jia, and I.~L. Moreno.
\newblock Voicefilter: Targeted voice separation by speaker-conditioned spectrogram masking.
\newblock \emph{arXiv preprint arXiv:1810.04826}, 2018.

\bibitem[Wang et~al.(2023)Wang, Cornell, Choi, Lee, Kim, and Watanabe]{wang2023tf}
Z.-Q. Wang, S.~Cornell, S.~Choi, Y.~Lee, B.-Y. Kim, and S.~Watanabe.
\newblock Tf-gridnet: Integrating full-and sub-band modeling for speech separation.
\newblock \emph{IEEE/ACM Transactions on Audio, Speech, and Language Processing}, 31:\penalty0 3221--3236, 2023.

\bibitem[Wu et~al.(2019)Wu, Xu, Zhang, Chen, Yu, Xie, and Yu]{wu2019time}
J.~Wu, Y.~Xu, S.-X. Zhang, L.-W. Chen, M.~Yu, L.~Xie, and D.~Yu.
\newblock Time domain audio visual speech separation.
\newblock In \emph{2019 IEEE automatic speech recognition and understanding workshop (ASRU)}, pages 667--673. IEEE, 2019.

\bibitem[Zhang et~al.(2016)Zhang, Zhang, Li, and Qiao]{zhang2016joint}
K.~Zhang, Z.~Zhang, Z.~Li, and Y.~Qiao.
\newblock Joint face detection and alignment using multitask cascaded convolutional networks.
\newblock \emph{IEEE signal processing letters}, 23\penalty0 (10):\penalty0 1499--1503, 2016.

\bibitem[Zhang et~al.(2021)Zhang, Xu, Hao, and Xu]{zhang2021online}
P.~Zhang, J.~Xu, Y.~Hao, and B.~Xu.
\newblock Online audio-visual speech separation with generative adversarial training.
\newblock In \emph{Proceedings of the 2021 7th International Conference on Computing and Artificial Intelligence}, pages 379--385, 2021.

\end{thebibliography}

\end{document}